\documentstyle{sup}
\vspace{-8pt}

\title[Superlattices and Microstructures, Vol.\ 20, No.\ 1, 1996]
{Long-range coherence and mesoscopic transport in N-S metallic structures}

\author[Superlattices and Microstructures, Vol.\ 20, No.\ 1, 1996]{Herv\'e Courtois, Philippe Gandit and Bernard Pannetier%
\cr\vspace{10pt}%
{\normalsize\it C.R.T.B.T.-C.N.R.S. in association with Universit\'e J. 
Fourier,}
\cr
{\normalsize\it 25 Av. des Martyrs, B.P. 166, 38042 Grenoble, 
France}\cr
Dominique Mailly%
\cr\vspace{10pt}%
{\normalsize\it L.M.M.-C.N.R.S., 196 Avenue H. Ravera, 92220 Bagneux, France}\cr
}

\pagerange{\pageref{firstpage}--\pageref{lastpage}}

\def\LaTeX{L\kern-.25em\raise.425ex\hbox{a}\kern-.075em\TeX}

\def\fakebold#1{\relax\ifvmode\leavevmode\fi%
\ifmmode%
\setbox0=\hbox{$#1$}%
\else%
\setbox0=\hbox{#1}%
\fi%
\kern-.02em\copy0 \kern-\wd0%
\kern .04em\copy0 \kern-\wd0%
\kern-.0125em\raise.02em\box0%
}%

\input epsf
\epsfverbosetrue
\begin{document}
\label{firstpage}
\maketitle
\sloppy
\begin{center}
\received{\today}
\end{center}
%

%
\begin{abstract}
We review the mesoscopic transport in a diffusive proximity superconductor made 
of a normal metal (N) in metallic contact with a superconductor (S). 
The Andreev reflection of electrons on the N-S interface is 
responsible for the diffusion of electron pairs in N. Superconducting-like 
properties are induced in the normal metal. In particular, the conductivity of the
N metal is locally enhanced by the proximity effect. A re-entrance of the 
metallic conductance occurs when all the energies involved (e.g. temperature 
and voltage) are small. The relevant characteristic energy is the 
Thouless energy which is $\hbar$ divided by the diffusion time for an 
electron travelling throughout the sample. In loop-shaped devices, a
$1/T$ temperature-dependent oscillation of the magnetoresistance arises 
with a large amplitude from the long-range coherence of low-energy pairs.
\end{abstract}

\section{Introduction}
The ability to fabricate small samples, thanks to the progress of the technology 
developped for the microelectronic industry, has opened a new field of research 
for the physics community. This field of research has been called mesoscopic 
physics since although the size of these systems is far bigger than the atomic 
scale, quantum mechanics is needed in order to understand their behavior. A new 
length scale has then been introduced, intermediate between the atomic scale and the 
macroscopic scale, namely the phase-breaking length $L_\varphi$ \cite{Curacao,Imry}. 
In pure metals at low temperature, $L_\varphi$ can reach several 
micrometers \cite{Pannetier-Rammal,Mohanty}.

The best known 
examples of mesoscopic physics are the universal conductance fluctuations and 
the weak localization. The first effect arises from the interference of 
the one-electron wave-function using all the possible paths to travel through the sample. 
This interference gives rise to magnetoconductance fluctuations, 
whose amplitude $\Delta G=e^2/h$ has the remarkable property to be universal, i.e. independent of 
the sample conductance. In a loop geometry, this interference 
gives rise to the well known Aharonov-Bohm oscillations with a flux-periodicity 
$2\phi_{0}=h/e$ \cite{Washburn}.

In cylinder or arrays geometries with a total length exceeding $L_\varphi$, Aharonov-Bohm 
oscillations and universal conductance fluctuations average to zero. But at low 
magnetic field, a special class of event survives to ensemble averaging. When one 
considers a given path and its time reversed dual, the wave-functions 
along the two paths interfer 
constructively if time reversal symmetry is conserved \cite{AAS}. In that 
special case, interferences increase the probability to return to the 
origin of a loop, and one observes a decrease of the conductance. For the 
loop geometry, the flux periodicity is $\phi_{0}=h/2e$ since the loop is 
travelled twice by the electron \cite{Sharvin}. 

Both conductance fluctuations and weak localization are mesoscopic 
effects in the sense that they appear on the length scale of the 
phase-breaking length $L_{\varphi}$. Their amplitude corresponds 
to the conductance ($e^2/h$) of one transport channel, even if many channels 
are involved. This comes from the fact that the different channels do 
not have a common phase reference. In the usual case of a sample 
made of an evaporated thin metallic film, the amplitudes of conductance 
fluctuations and weak localization are small compared to the total conductance.

The mesoscopic effects are not intrinsically limited to a small 
amplitude. The most beautiful example is the "mesoscopic 
superconductivity" arising in a normal metal N from the Andreev 
reflection of electrons at a superconductor S. The 
superconductivity in S results in a macroscopic phase-coherent pair 
condensate. In these N-S systems, the coupling of normal 
electrons to the pair condensate induces coherence effects which add 
constructively and reach a total large amplitude.

In this review paper, we will discuss the transport properties of 
devices made of a normal metal in metallic contact with a superconductor. 
This scientific area has met a remarkable new interest in 
the recent years, due to novel experiments on samples with new mesoscopic 
designs \cite{Revue_Lambert,Rev_Beenakker,Comments}. Here, we will restrict 
to the case of transport in diffusive metals 
in the presence of Andreev reflection. The Josephson effect 
\cite{Exp-Joseph,Th-Joseph} will not be discussed as we will 
concentrate on non-equilibrium properties. We will focus on the 
regime where the resistance of the N-S interface is negligeable compared to the 
metallic resistance of the wire.

\section{An introduction to the quasiclassical theory of the proximity 
effect}

\subsection{The Andreev reflection}

Let us consider a sample made of a normal metal in contact with a 
superconductor. At a temperature well below the superconducting transition of 
S and at low bias voltage, the thermal energy $k_B T$ and the electrostatic 
energy $eV$ are much smaller than the gap of S. As a consequence, normal electrons 
arriving on the N-S interface cannot find any available states in S 
and are Andreev-reflected \cite{Andreev}.

\begin{figure}
\epsfxsize=13 cm \epsfbox{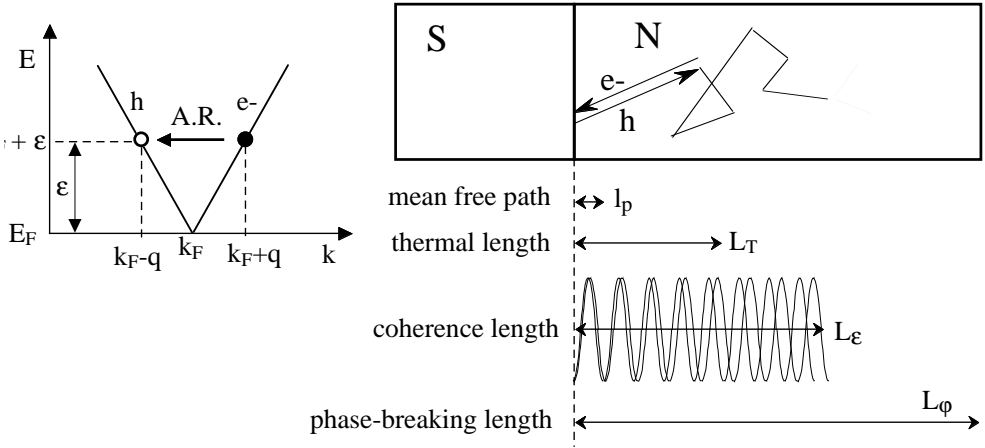}
\caption{Left : Schematic of the Andreev reflection process. An
electron with a small extra energy $\epsilon$ compared to the 
Fermi energy $E_F$ arrives from the N side on the N-S interface. There 
is a wave-vector mismatch of $2q$ between 
the incident electron and the reflected hole. Right : Relevant length 
scales with their schematic respective amplitude in a metallic thin 
film. The energy-dependent coherence length $L_\epsilon$ is the length 
over which the two components of the electron pair acquire a phase 
difference of order $\pi$.}
\label{Andreev}
\end{figure}

In the Andreev reflection, the electron 
is reflected as a hole, while a Cooper pair is transmitted in the superconductor,
see Fig. \ref{Andreev}. The spin and every component of the electron momentum 
are reversed. In this scope, a N-S interface is similar to a phase-conjugating 
mirror in optics. As it has the opposite momentum, the reflected hole traces 
back the path of the incident electron. The phase acquired by the electron is 
"eaten up" as the hole retraces back the trajectory of the incident 
electron.

The electron and the hole do not form a pair 
since they are not present at the same time in the metal, but the 
Andreev reflection correlates two electron states in N. The reflection of the 
hole may also be seen as the absorption of an electron by the 
superconductor. In this scope, two electrons travel on the same path 
through the N metal and are absorbed as a Cooper pair in S. This effect can be 
reversed, so that an electron pair from S can diffuse in the N metal. In 
the following, we will indifferently use both languages : reflection of an 
electron into a hole (and vive-versa) or diffusion of an electron pair in N.

If one looks more precisely, the reversal of the electron momentum 
is perfect only at the Fermi level. Let us consider an electron with a 
small extra energy $\epsilon$ with respect to the Fermi level energy $E_F$. Its 
wave-vector $k_e = k_F + q$ slightly differs from the Fermi 
wave-vector $k_F$. The reflected hole has a slightly different 
wave-vector $k_h = k_F - q$, see Fig \ref{Andreev}. The incident electron 
and the reflected hole have a difference $\delta k$ in the 
perpendicular component of the wave-vector so that :
\begin{equation}
\delta k= 2q = k_F \frac{\epsilon }{E_F}.
\end{equation}

Let us now focus on the case where the normal metal is in the dirty limit. 
This limit is valid when the coherence length of electron pairs 
(defined below) is 
much larger than the elastic mean free path. We will also assume  
that the phase-breaking length $L_{\varphi}$ is much larger than the 
sample length $L$. By diffusing over a distance $L$ in N, the phase 
difference between the two electrons of the diffusing pair increases 
monotonically. After diffusion during a time $t$ over a distance 
$L \simeq \sqrt{D t}$ from the interface, the phase shift is :
\begin{equation}
\delta \varphi = \frac{L^2 }{ L_\epsilon^2}, 
\label{Dephasage}
\end{equation}
where we introduce the energy-dependent coherence length :
\begin{equation}
L_\epsilon = \sqrt{\frac{\hbar D}{ \epsilon}},
\end{equation}
$D$ being the diffusion coefficient in N. The Eq. \ref{Dephasage} 
means that the phase drift between the two components of the 
diffusing pair will be small as long as $L$ is small compared to 
$L_\epsilon$. In this regime, the two electrons are scattered in the 
same way by impurities and their trajectories are quite the same. 
As a consequence, the two electrons appear as linked in pair.

At a distance of the order of $L_{\epsilon}$, the phase 
difference is non-negligeable. At the same time, the trajectories of 
the electron and the hole are separated by a distance or order of the 
Fermi wavelength \cite{Blom}. Further scattering of the two particles 
will therefore be decorrelated and the pair will break apart. For this reason, we 
call $L_{\epsilon}$ the coherence length of the electron pair.

It is remarkable that the coherence length $L_\epsilon$ coincides 
with the thermal length $L_T$ at the energy $ \epsilon = 2\pi k_B T$. 
The length $L_T$ is usually given in the textbooks \cite{deGennes} as 
the characteristic decay length for the proximity effect. The thermal 
length $L_{T}$ is characteristic of a thermal equilibrium electron 
distribution while the coherence length $L_{\epsilon}$ is 
characteristic of electrons at a given energy. At the Fermi level, 
the coherence length $L_\epsilon$ is infinite, independently of 
temperature. The actual coherence length is limited by the 
phase-breaking length $L_{\varphi}$ of a single electron. In 
conclusion, the effective coherence length of an electron pair in N varies from about 
the thermal coherence length $L_T$ (at $\epsilon= 2 \pi k_{B}T$) to the 
phase-coherence length $L_\varphi$ at low energy (at $\epsilon = 0$).

The phase drift between the electron and the reflected hole can also 
be written as :
\begin{equation}
\delta \varphi = \frac{\epsilon}{\epsilon_c}
\label{Thouless}
\end{equation}
where
\begin{equation}
\epsilon_c = \frac{\hbar D }{ L^2}
\end{equation}
is the Thouless energy of a sample of 
length $L$. The interpretation of Eq. \ref{Thouless} is again very simple. 
At a given distance $L$, only electrons with an energy below the 
Thouless energy $\epsilon_{c}$ are still correlated in pairs.

\subsection{The Usadel equations}

The diffusion of superconductivity in any inhomogenous structure can 
be described by the Gorkov Green's functions. In the limit where all 
the experimentally-relevant length scales are much larger than the 
Fermi wavelength, the simplification of this fully quantum theory 
into the quasiclassical theory is valid. In the diffusive 
regime where the elastic mean free path is small, the Usadel 
equations \cite{Usadel} are obtained. In this 
framework, the weak localization effects and the conductance 
fluctuations are neglected.

The quasiclassical theory has been developped in its full integrity by 
many groups \cite{Volkov,Nazarov96,Volkov-Lambert,Zhou,Wilhelm,Yip}. 
A review of the applications of this theory to the field of 
mesoscopic superconductivity is included in this volume 
\cite{Revue_Belzig}.
Here we would like to present a simplified version that keeps the 
essential physical features of the original theory. For instance, it is 
very convenient to linearize the Usadel equation. This is acceptable 
in most of the experimental situations as far as the distance from 
the interface is not too small compared to the thermal length $L_T$ 
\cite{JLTP}. 
In this case, the Usadel equations reduce to an equation for the pair 
amplitude $F$ in the normal metal :
\begin{equation} 
\hbar D \, \partial^{2}_{x} F + \{\, 2 i\epsilon - \frac{\hbar 
D}{L_{\varphi}^2}\, \} F 
= 0
\label{diffusion}
\end{equation}
The pair amplitude $F$ is a two-variables function of the position 
$x$ and the energy $\epsilon$. We neglected here the pairing 
interaction in N and the inelastic scattering. The Eq. \ref{diffusion} can be 
intuitively understood as a diffusion equation for the pair 
amplitude. From the equation, one can derive the following 
facts : (i) the natural length scale of the pair 
amplitude diffusion is $L_\epsilon$ ; (ii) the diffusion is bound by an 
exponential cut-off at the distance $L_\varphi$. These features 
confirm our previous qualitative discussion.

\begin{figure}
\epsfxsize=12 cm \epsfbox{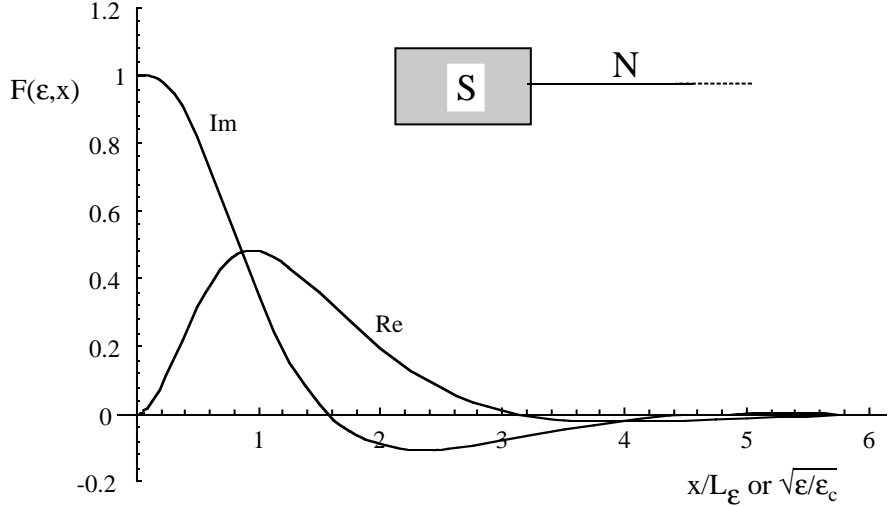}
\caption{Plot of the imaginary and real part of the pair amplitude 
wave-function $F(\epsilon,x)$ calculated in the framework of the linearized 
Usadel equation. The horizontal axis can be read either as the distance $x$ 
normalized to the energy-dependent coherence length $L_\epsilon$ 
(here the energy $\epsilon$ is fixed) or as the square root of the energy 
$\epsilon$ in units of the local energy scale $\epsilon_x = \hbar D/x^2$ 
related to the distance $x$ (here the distance $x$ is fixed). Inset : 
the considered sample geometry made of an unfinite N wire in contact 
with a superconductor S.}
\label{Usadel}
\end{figure}

As an exemple, let us consider the case of a quasi-1D sample made of a long 
normal metal wire in perfect contact with a superconducting island, see 
Fig. \ref{Usadel} inset. The phase-breaking length $L_{\varphi}$ is assumed to be 
infinite. Solving the Usadel equation, one can obtain the 
pair amplitude as a complex function $F(\epsilon,x)$ of both the 
distance x from the S interface and the energy $\epsilon$.
Fig. \ref{Usadel} shows the analytically-calculated real and imaginary 
parts of the pair amplitude. The units of the horizontal axis can be 
read either as the distance x from the S interface in units of the 
coherence length $L_\epsilon$, or as the square root of the energy 
$\epsilon$ above the Fermi level in units of the local energy 
scale $\epsilon_x = \hbar D / x^2$ related to the distance $x$. This 
energy scale coincides with the Thouless energy for $x=L$.

As any of the two variables (energy or position) increases, the imaginary 
part of $F(\epsilon,x)$ decays in an oscillating way, see Fig. \ref{Usadel}.
This is true when one fixes the energy $\epsilon$ and let the distance $x$ 
from the S interface grow, or if one fixes the distance $x$ and let the 
energy $\epsilon$ grow. The energy scale and length scales of decay are 
respectively $\epsilon_x =\hbar D/ x^2$ and $L_\epsilon = \sqrt{\hbar 
D /\epsilon}$. The measurement of the local density of states 
\cite{Gueron,Dos_sns} or of the magnetic screening \cite{Mota} gives a direct 
experimental access to the imaginary part of $F$.

The proximity effect on dissipative transport is connected to the real 
part of the pair amplitude. From Fig. \ref{Usadel}, the real part is zero 
at the S interface and/or at zero energy. At fixed distance and variable 
energy, it is maximum at the energy $\hbar D/ x^2$. At a fixed energy, the real 
part is maximum at the distance $L_\epsilon$ from the interface. At high energy, 
the real part also decays in an oscillating way. As we will see in the 
following, the peak at energy $\epsilon \simeq \epsilon_c$ in the real part of $F$ is 
responsible for the re-entrance effect.

\subsection{The spectral conductance}

In the quasiclassical theory, the coherent transport can be described 
by a local conductivity. The proximity effect results in a local conductivity 
enhancement $\delta \sigma$ which is connected to the real part of the pair 
amplitude $F$ :
\begin{equation} 
\delta \sigma(\epsilon,x) = \sigma_N (Re [F(\epsilon,x)])^{2}
\label{conductance}
\end{equation}
for small $F$, $\sigma_N$ being the normal-state conductivity. Let us note 
that this excess conductivity is energy-dependent and always 
positive : the conductance will always increase due to the 
proximity effect.

Coming back to the sample geometry of Fig. \ref{Usadel}, we 
can draw some conclusions concerning the energy dependence of the sample
conductivity. Starting from the zero-energy where the normal-state conductivity 
is recovered, the conductivity is enhanced by about $25 \%$ at the maximum 
located at $\epsilon \simeq \hbar D/x^2$. At higher energy, 
the excess conductivity decay as $1/\sqrt{\epsilon}$. The most striking 
prediction is that the normal-state conductivity is recovered at zero 
energy. It is the origin of the re-entrance effect.

The spectral conductance $g(\epsilon)$ is defined as the conductance for 
electrons with a given energy $\epsilon$ above the Fermi level flowing through 
the sample. It can be calculated using the classical circuit theory. In a quasi 
1D geometry, it is equal to the wire section S divided by the integral of 
the local resistivity over the sample length $L$ :
\begin{equation}
g(\epsilon)=S [\int_{0}^{L}{ \frac{1}{\sigma_{N}+\delta \sigma(\epsilon,x)}dx}]^{-1}
\end{equation}
The spectral conductance is experimentally accessible 
as it coincides with the zero-temperature limit of the differential conductance at 
finite bias :
\begin{equation}
G(V=\epsilon/e,T=0)=g(\epsilon)
\end{equation}
At finite temperature and zero bias, the measured conductance 
is essentially an integration over a thermal window of width $k_{B}T$ :
\begin{equation}
G(V=0,T) = \int_{-\infty}^{\infty} g(\epsilon ) [4k_B T \,{cosh}^2(\epsilon/2k_B T)]^{-1}  d\epsilon 
\label {total}
\end {equation}

As an example, let us consider the case of a N metalic wire of length $L$ 
inserted between a N reservoir and S island. The spectral conductance 
has a maximum of $15 \%$ of the normal-state conductance at an energy 
of about 5 times the Thouless energy $\hbar D/ L^2$ \cite{Wilhelm}. 
At zero energy, the spectral conductance is equal to the normal-state 
conductance. This effect has been called the re-entrance effect 
since, coming from the finite energy or temperature regime, 
there is a re-appearance of the plain metallic conductance at zero 
energy. This result can be exactly derived from the full Usadel equations 
\cite{Volkov,Nazarov96,Volkov-Lambert,Wilhelm,Yip}. It is also in 
agreement with the prediction of the random matrix theory 
\cite{Beenakker} and Bogoliubov-de Gennes equations 
\cite{Volkov-Lambert}.

\section{The re-entrance effect}

\subsection{The experimental observation}

In our study, we have focused on the case of noble metals like Cu
evaporated in thin films. In these systems, the 
elastic mean free path $l_e \simeq \, 10 nm$ is quite small. The 
Thouless energy is about $2 \, \mu eV \simeq 20 \, mK$ for a sample length 
of $1 \,\mu m$. The thermal length $L_T$ is of the order of $100 \, nm$ at 1K. The 
phase-breaking length $L_\varphi \simeq 1 \mu m$ is approximately constant in our 
temperature range ($T < 1 \, K$). Let us note that the elastic mean free path and 
phase-breaking length scales are quite different in evaporated thin 
metallic films. This is a major advantage of noble metals compared to 
doped semiconductors, because one can easily distinguish which length scale 
is related to a given physical effect.

\begin{figure}
\epsfxsize=15 cm \epsfbox{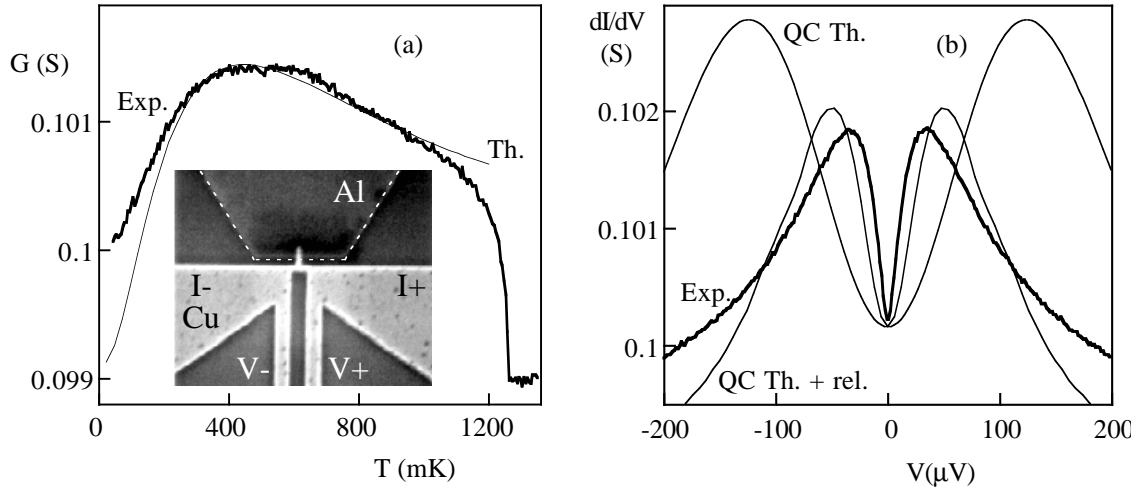}
\caption{(a) : Inset shows the micrograph of a typical sample. A 
dotted white line which follows the edges of the Al island has been drawn 
for clarity. Main figure : temperature dependence of the conductance 
measured between the two Cu reservoirs. The experimental curve (Exp.) 
is shown in parallel with the theoretical curves (Th.). The theoretical 
curves have been calculated with the non-linear Usadel equation taking 
into account the finite gap of the superconductor. The phase-breaking length is 
considered as infinite. (b) Voltage dependence of the differential 
conductance of the same sample. Compared to the plain theoretical 
curve (QC Th.), the theoretical curve labeled (QC Th. + 
rel) includes the additional effect of relaxation in the reservoirs.}
\label{Reentrance}
\end{figure}

Fig. \ref{Reentrance}a inset shows the micrograph of the sample 
we will consider \cite{LT}. This T-shaped sample 
is made of three Cu arms joining three wide pads : two are made of 
Cu, one is of Al. Thin films of Al are superconducting below about 1.3 K. 
The Cu-Al interface has been carefully prepared in order to
obtain a highly transparent interface. Each of the pads is connected to 
a voltage and a current probe. We will present measurement 
of the conductance measured 
between the two Cu reservoirs. With this geometry, we can assert that 
the current redistribution in the vicinity of the Cu-Al 
interface due to the superconductivity of Al \cite{NS_a_2D} has a 
negligeable effect on the measured conductance.

Fig. \ref{Reentrance} shows the temperature dependence of the 
conductance measured with a low bias voltage. As the temperature 
decreases below the superconducting transition of Al, the conductance 
first rises. It afterwards reaches a maximum and eventually decreases 
down to the lowest temperature. The same kind of
behaviour can be seen in the voltage dependence of the differential 
conductance. The differential conductance exhibits a minimum at zero bias, 
a maximum at a finite bias voltage and then a decrease at 
large bias.

This experiment provides the proof for the re-entrance of the 
metallic conductance in a mesoscopic proximity superconductor. A 
similar effect has been observed in other metallic samples \cite{Petr_1998} 
and semiconductor-superconductor structures, the semiconductor 
being a two-dimensional electron gas \cite{Hartog}. Here, we have 
been able to track the re-entrance effect as a function of both 
temperature and voltage and, as discussed in the following, to 
describe quantitatively our data with the theory.

\subsection{Comparison with the theory}

We used the quasiclassical theory to quantitatively describe the 
experimental results. The full non-linear Usadel equations were solved 
numerically in the precise sample geometry. In both Fig. 
\ref{Reentrance}a and b, calculated 
curves are shown in comparison of the experimental one. The physical 
parameters used in the calculation match the experimental values and 
are the same in $G(T)$ and $dI/dV(V)$ data. The Thouless energy used 
in the calculation is $\epsilon_{c} = 15.5 \, \mu eV$ and corresponds to 
a temperature $\epsilon_{c}/ k_{B}=180 \, mK$. 
Due to the geometry of the sample, the maximum of the spectral 
conductance is expected at $3.8 \, \epsilon_{c}$. We took into account 
the finite amplitude of the gap in S, as it is not much larger than 
the sample Thouless energy.

The temperature dependence data are well fitted by the theoretical calculation. 
The main difference is at zero energy and zero temperature, where one 
expects the normal-state value of the conductance. This is not exactly 
the case in the experiment. This may be an effect of the interactions 
in the normal metal \cite{Nazarov96,JLTP}.

In the voltage dependence data, there is a large discrepancy between the experimental and 
the calculated curves. We found it was necessary to take into account the energy 
relaxation in the reservoirs \cite{JLTP} : if heating of the reservoirs due 
to the bias current is included, one recovers a fairly good agreement.

\subsection{Discussion}

When one integrates the spectral conductance for a thermal 
distribution of electrons, one finds that the conductance change is 
analog to the effect of superconductivity over a length $L_T$ from 
the interface \cite{Zhou}. The total conductance change then varies 
as $1/\sqrt{T}$ with temperature $T$. This does not mean that this 
region exhibits full superconducting properties. The re-entrance effect 
brings the counter-proof of it, since at zero temperature and zero bias, 
the proximity effect on conductance vanishes.

From the theoretical point of view, it is clear that the peak of the real part 
of $F(\epsilon,x)$ at finite energy (see Fig. \ref{Usadel}) describes 
perfectly the re-entrance effect. A more physical explanation of the 
re-entrance effect is not an easy task. If one looks at the behaviour of 
the conductivity at a given energy throughout the sample, the conductivity 
is the normal-state one both near the interface and far away 
in the sample. The peak appears in the vicinity of the distance 
$L_\epsilon$ from the S interface. Strikingly, it is the point {\it 
where the electron pairs break}. Everything happens as if it were the 
breaking of the pairs into two free electrons which is responsible 
for the conductivity enhancement.

\section{Long-range coherence}

\subsection{An interference experiment}

\begin{figure}
\epsfxsize=15 cm \epsfbox{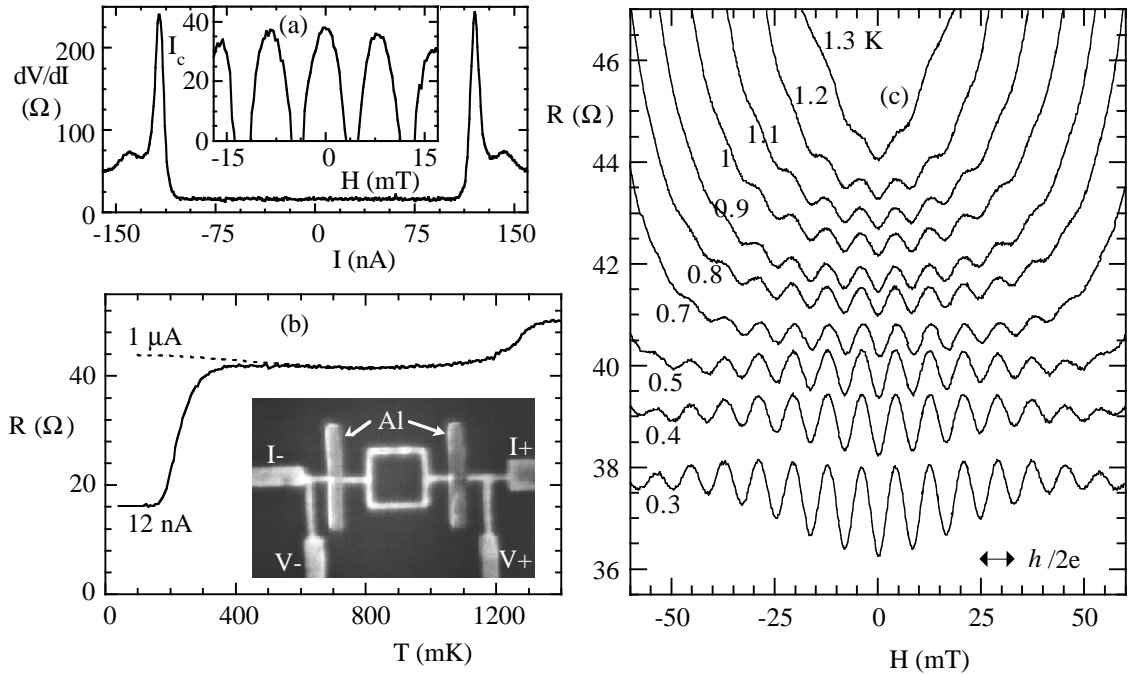}
\caption{(a) Sample differential resistance dependence on the bias 
current at 40 mK. Inset shows the critical current (in nA) measured 
at 150 mK with a $25 \Omega$ differential resistance criteria as a 
function of the magnetic field. Flux 
periodicity of the critical current oscillations is $h/2e$. (b) 
Temperature dependence of the sample resistance with a measurement 
current 12 nA and 1 $\mu A$. Inset : Micrograph of a typical sample 
made of a Cu square loop with 4-wire measurement contacts, in contact 
with two Al islands (vertical). Size of the loop is 500 X 500 $nm^2$, width 
50 nm, thickness 25 nm. Centre-to-centre distance between the 150 nm 
wide Al islands is 1 $\mu m$. The length L of the N part of the S-N-S 
junction is 1.35 $\mu m$. The elastic mean free path is 16 nm, the 
thermal coherence length is 99 nm at 1 K and the phase-breaking 
length is about 1.9 $\mu m$. (c) : Low-field magnetoresistance for T 
= 0.3; 0.4; 0.5; 0.7; 0.8; 0.9; 1; 1.1; 1.2 and 1.3 K. Curves have 
been arbitrarily shifted for clarity. 
Oscillations of periodicity $h/2e$ and amplitude $e^2/h$ at T = 0.8 K 
are visible.}
\label{RTetRHboucle}
\end{figure}

Interference experiments are very useful to distinguish the small 
coherent contribution from a large background in a complex system. The 
Aharonov-Bohm geometry allows a direct control of the phase shift by 
an external magnetic flux. We 
fabricated a Cu loop-shaped sample which is locally in 
contact with two Al islands, see Fig. \ref{RTetRHboucle}b inset. The 
sample fabrication was such that the interface transparency is very 
high (see Ref. \cite{CourtoisPrl} for details). Let us point out that 
in all the described measurements, only the conductance of the N 
metal is measured since the bias current flows through the N wire 
only. There is no net current through the N-S junctions.

The experimental results are shown in Fig. \ref{RTetRHboucle}. When 
the temperature is decreased below the critical temperature $T_c$ of Al, 
the resistance of the Cu wire exhibits a sharp drop. This is partly due to the 
short-circuit of the length of Cu wire which is directly under the Al 
island, see Fig. \ref{RTetRHboucle}. At a temperature near 200 mK, 
the Josephson effect between the two Al islands appears as a sharp drop of the measured 
resistance. The residual resistance is due to the N wire in series between 
each voltage probe and the adjacent Al island. When one sweeps the magnetic 
field in the low temperature regime 
where the Josephson effect is present, a SQUID effect is observed in 
the critical current data (see Fig. \ref{RTetRHboucle}a inset). 

Let us note that the re-entrance effect is also observed in 
this experiment since the sample resistance increases with 
temperature in the cases where Josephson effect is negligeable 
(intermediate temperature or high current bias), see Fig. 
\ref{RTetRHboucle}b. The increase of resistance is related to 
the contribution of the lateral sides of the samples, outside the Al 
islands.

\subsection{The magneto-resistance oscillations}

In the intermediate temperature region $(T \geq 0.3 K)$, the normal metallic 
loop between the two Al islands is resistive. Large $h/2e$-periodic 
magnetoresistance oscillations are observed, see Fig. \ref{RTetRHboucle}c. 
This observation is consistent with previous experiments, including 
those reported by V. Petrashov et al. \cite{Petrashov} and A. Dimoulas et al. \cite{Dimoulas}.

Any influence of the Josephson current fluctuations \cite{MT} in the 
magnetoresistance oscillations has to be discarded since the 
Josephson coupling is vanishingly small in this temperature range. 
Moreover, the large amplitude ($50e^2/h$ at 1K) of the oscillations 
compared to the quantum conductance prevents any interpretation in terms 
of weak localization. The magnetoresistance oscillations are 
therefore definitely due to the proximity effect on the resistive 
transport in the normal metal, i.e. to the quantum interference of 
electron pairs.

In the considered high temperature range, the thermal coherence length 
$L_T$ is much smaller than the loop perimeter. Despite this, 
the proximity-induced conductance enhancement is modulated by a 
magnetic flux. This proves that the decay length of the proximity 
effect is much larger than the thermal length $L_T$. Actually, the 
final cut-off for the proximity effect is the phase-breaking length 
$L_{\varphi}$ \cite{Lindelof}.

\begin{figure}
\epsfxsize=15 cm \epsfbox{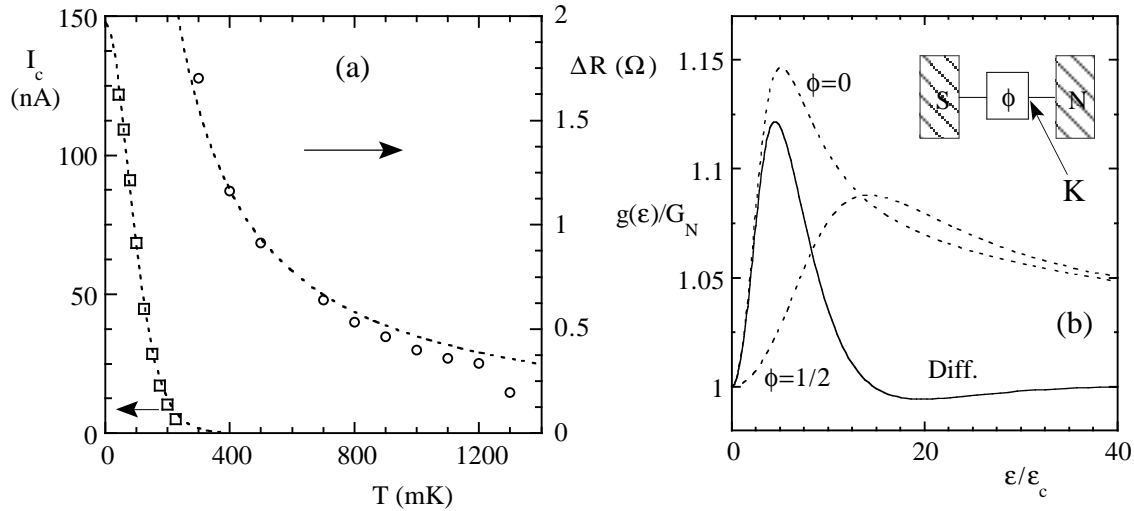}
\caption{(a) Left scale : Temperature dependence of the critical 
current derived from Fig. 4 data with a $25 \Omega$ differential 
resistance criteria. Dashed line is a guide to the eye. Right scale : 
Temperature dependence of the amplitude of the magnetoresistance 
oscillations, Dashed line is a 1/T power-law fit. (b) Sketch of the 
spectral conductance as a function of the voltage bias in a 
loop-shaped sample with zero flux and half a flux quantum in the loop 
(dotted lines). The ratio distance S-K over distance S-N is 0.6. The 
full curve gives the difference of spectral conductance between 
these two cases. Inset shows the sample model. The arrow shows the 
point K where the pair amplitude is zero when the flux through 
the loop is $\varphi=\phi_{0}/2$.}
\label{ICetMR}
\end{figure}

\subsection{Analysis of the oscillations amplitude}

From the Fig. \ref{RTetRHboucle}b, we observe that the 
magnetoresistance oscillations are very robust against the 
temperature. The oscillations at 1.3 K have an amplitude which is 
barely reduced as compared to the data at 0.7 K. Fig. \ref{ICetMR} 
shows the temperature dependence of both the magnetoresistance 
oscillations amplitude and the Josephson critical current. The 
difference in behaviour is striking. While the critical current is 
exponentially suppressed at high temperature $T$, the magnetoresistance 
oscillations amplitude follows a $1/T$ power law. The deviation at 
high temperature is related to the closure of the gap near the 
superconducting transition of Al.

The $1/T$ power-law can be explained in a intuitive way, as discussed 
in the following. The magnetic flux in the loop creates interference between electron 
pairs travelling along one side of the loop or the other side. Let us 
consider the contribution of only one Al island. With two islands, the effects will add up.

Let us restrict to the two extreme cases of an 
integer or integer plus one half number of flux quantum in the loop. In 
the integer plus one half case, there is a destructive interference ($F = 0$) of the 
diffusing pairs at the node K opposite to the S island, see Fig. 
\ref{ICetMR}b. In the integer case, the magnetic flux has no effect on the pair 
amplitude. In summary, the point K acts as an effective 
N reservoir when half a flux quantum threads the loop. This reduces 
the effective sample length from $L$ to a given fraction of $L$.

Fig. \ref{ICetMR}b shows the calculated spectral conductance in the 
two extreme cases $\varphi = 0$ and $\varphi=\phi_{0}/2$. The difference 
between the two spectral conductances is restricted at low energy, 
since at point K, only electrons with an energy close to the Fermi 
level may remain coherent. The pairs with a large energy are 
unaffected by the flux, as their coherence length $L_{\epsilon}$ is much smaller 
than the loop perimeter. This makes the flux an unique tool for 
selecting the low-energy, long-range coherence in N-S structures with 
an Aharonov-Bohm loop.

Let us now estimate the impact of the interference on 
the sample conductance at finite temperature. We assume that the 
temperature $T$ is much larger than 
the Thouless temperature $\epsilon_c / k_B$. The effect of the 
interference is limited to an energy window of about $10 \epsilon_c$, 
see Fig. \ref{ICetMR}b. In a 
first approximation, the spectral conductance enhancement (of about $15 \%$) 
is cancelled for this population. The relative amplitude of the 
magnetoresistance oscillation is then given by $15 \%$ of the ratio 
of this population ($10 \, \epsilon_c$) over the thermal 
population ($\epsilon \leq k_B T$). This gives a ratio $1.5 \, \epsilon_c / 
k_B T$.

This qualitative argument is confirmed by a 
rigourous calculation based on the Usadel equations \cite{Wilhelm}. A ratio 
of about $\epsilon_c / k_B T$ is in good agreement with the 
experiment. It has a natural interpretation, since it is the 
fraction $\epsilon_{c}$ of electrons from the thermal distribution 
$k_{B}T$ which remain 
coherent over the normal metal loop. The Thouless energy appears here again 
as the natural energy scale for the proximity superconductivity.

\section{Conclusion and Perspectives}

Although it used to appear as old-fashioned, the proximity effect has 
recently known a noticeable revival. This is due to the discovery of 
its natural mesoscopic status and the prodigeous progresses in 
nanofabrication. A major 
example is its natural length scale being the phase-breaking length 
of individual electrons $L_\varphi$. The proximity superconductivity in 
mesoscopic samples shows many unique features in the field of 
mesoscopic physics. It is an ensemble-averaged effect which affects 
at order 0 in $k_{F}l_{p}$ the electronic conductance of metallic 
samples \cite{Zhou}. The 
Thouless energy $\epsilon_c$ is the relevant energy scale for the 
proximity effect. This energy appears in the Josephson effect, the 
re-entrance effect, the density of states and the magneto-conductance 
oscillations. As an illustration, let us note that the Josephson coupling 
will appear in the S-N-S sample at the same temperature as 
the normal-state conductance would re-enter in the S-N-N sample.
 
Some experimental works have lead to discussions on the relevance of 
the theory, but up to now the theory has eventually explained the 
main experimental results. For instance, the resistance enhancement 
observed in some N-S structures \cite{Petr_Anomaly} is now reliably 
understood as a 2D effect of redistribution of current lines as the 
temperature is decreased below the S metal superconducting transition 
\cite{NS_a_2D,Lambert_2D}. An open question consists of the interplay between the Andreev 
reflection and weak localization diagrams which could lead to the observation of 
$h/4e$-periodic oscillations of the magnetoresistance \cite{Spivak82}. This effect has not 
been observed yet since the only observation of $h/4e$ periodicity 
\cite{Petrashov} was in fact due to the tricky geometry of the 
samples \cite{Zaitsev}.

One of the challenging perspectives in the field is the effect of interactions 
in the normal metal. This motivates the study of proximity effect in exotic 
materials like ferromagnetic metals \cite{FS}.

\section{Acknowledgements}

We are grateful to P. Charlat who performed most of the experimental work on 
the re-entrance effect. We thank F. Wilhelm, F. Zhou, B. Spivak, T. Martin, 
A. Zaikin, and A. F. Volkov for illuminating discussions on the 
theory. We benefited of collaborations with M. Giroud and K. Hasselbach.
We acknowledge financial support from R\'egion Rh\^one-Alpes, 
D.R.E.T. and European Union through TMR contract FMRX-CT97-0143 "Dynamics of 
Superconducting Nanocircuits".


\label{lastpage}

\end{document}